\pdfoutput=1
\documentclass[preprint,aps,superscriptaddress]{revtex4}
\usepackage{graphicx,amssymb,bm,natbib,amsfonts,amsmath}
\usepackage{hyperref}
\usepackage{epstopdf}
\usepackage{color}

\newcommand {\C}{\textcolor {blue}}

\begin{document}

\title{Glass-like recovery of antiferromagnetic spin ordering in a photo-excited manganite  Pr$_{0.7}$Ca$_{0.3}$MnO$_3$}

\author{S.Y. Zhou}
\altaffiliation{Correspondence should be sent to syzhou@mail.tsinghua.edu.cn, YChuang@lbl.gov and RWSchoenlein@lbl.gov.}
\affiliation{Materials Sciences Division, Lawrence Berkeley National Laboratory, Berkeley, CA 94720, USA}
\affiliation{State Key Laboratory of Low Dimensional Quantum Physics and Department of Physics, Tsinghua University, Beijing 100084, China}
\affiliation{Advanced Light Source, Lawrence Berkeley National Laboratory, Berkeley, CA 94720, USA}
\author{M.C. Langner}
\affiliation{Materials Sciences Division, Lawrence Berkeley National Laboratory, Berkeley, CA 94720, USA}
\author{Y. Zhu}
\affiliation{Materials Sciences Division, Lawrence Berkeley National Laboratory, Berkeley, CA 94720, USA}
\affiliation{Department of Applied Science, University of California, Davis, CA 95616,USA}
\author{Y.-D. Chuang}
\altaffiliation{Correspondence should be sent to syzhou@mail.tsinghua.edu.cn, YChuang@lbl.gov and RWSchoenlein@lbl.gov.}
\affiliation{Advanced Light Source, Lawrence Berkeley National Laboratory, Berkeley, CA 94720, USA}
\author{M. Rini}
\affiliation{Materials Sciences Division, Lawrence Berkeley National Laboratory, Berkeley, CA 94720, USA}
\author{T.E. Glover}
\affiliation{Advanced Light Source, Lawrence Berkeley National Laboratory, Berkeley, CA 94720, USA}
\author{M.P. Hertlein}
\affiliation{Advanced Light Source, Lawrence Berkeley National Laboratory, Berkeley, CA 94720, USA}
\author{A.G. Cruz Gonzalez}
\affiliation{Advanced Light Source, Lawrence Berkeley National Laboratory, Berkeley, CA 94720, USA}
\author{N. Tahir}
\affiliation{Advanced Light Source, Lawrence Berkeley National Laboratory, Berkeley, CA 94720, USA}
\affiliation{National Center for Physics, Islamabad, Pakistan}
\author{Y. Tomioka}
\affiliation{Nanoelectronics Research Institute, National Institute of Advanced Industrial Science and Technology (AIST) Tsukuba Central 4, 1-1-1 Higashi Tsukuba 305-8562, Japan}
\author{Y. Tokura}
\affiliation{Department of Applied Physics, University of Tokyo, Bunkyo-ku, Tokyo 113-8656, Japan}
\affiliation{Cross-Correlated Materials Research Group (CMRG) and Correlated Electron Research Group (CERG), Advanced Science Institute, RIKEN, Wako 351-0198, Japan}
\author{Z. Hussain}
\affiliation{Advanced Light Source, Lawrence Berkeley National Laboratory, Berkeley, CA 94720, USA}
\author{R.W. Schoenlein}
\altaffiliation{Correspondence should be sent to syzhou@mail.tsinghua.edu.cn, YChuang@lbl.gov and RWSchoenlein@lbl.gov.}
\affiliation{Materials Sciences Division, Lawrence Berkeley National Laboratory, Berkeley, CA 94720, USA}

\date{\today}

\begin{abstract}
{\bf Electronic orderings of charges, orbitals and spins are observed in many strongly correlated electron materials, and revealing their dynamics is a critical step toward understanding the underlying physics of important emergent phenomena. Here we use time-resolved resonant soft x-ray scattering spectroscopy to probe the dynamics of antiferromagnetic spin ordering in the manganite Pr$_{0.7}$Ca$_{0.3}$MnO$_3$ following ultrafast photo-exitation. Our studies reveal a glass-like recovery of the spin ordering and a crossover in the dimensionality of the restoring interaction from quasi-1D at low pump fluence to 3D at high pump fluence. This behavior arises from the metastable state created by photo-excitation, a state characterized by spin disordered metallic droplets within the larger charge- and spin-ordered insulating domains. Comparison with time-resolved resistivity measurements suggests that the collapse of spin ordering is correlated with the insulator-to-metal transition, but the recovery of the insulating phase does not depend on the re-establishment of the spin ordering. }

\end{abstract}

\maketitle

Nanoscale electronic orderings of charges, orbitals and spins (e.g. into stripe or checkerboard patterns) exist in many strongly correlated electron materials, and they are directly relevant to emergent phenomena, such as colossal magnetoresistance (CMR) \cite{DagottoRP, TokuraRPP06}, high temperature superconductivity \cite{Tranquada}, and multiferroic behavior \cite{LuFe2O4}. Elucidating the fundamental origin of these ordered phases and their dynamic interplay remain important scientific challenges. Transient photo-excitation can drive transitions between competing states, e.g. photo-induced transient insulator-metal transition (IMT) in charge/orbital/spin ordered CMR manganites \cite{FiebigSci,RiniNat}, and transient superconductivity in a stripe-ordered cuprate \cite{CavalleriSci}. This is also an effective means for separating the strong coupling between electron, lattice, orbital and spin degrees of freedom based on their disparate time responses. Moreover, transient excitations can be used to create metastable phases from which the re-establishment of the ordered states may be directly observed via ultrafast probes. The application of resonant soft x-ray scattering spectroscopy (RSXS) as an ultrafast time-resolved probe (TR-RSXS)  \cite{Durr,LSMO,CuO,LeeWS,Forst} fills a critical knowledge gap on the electronic ordering dynamics by providing direct spectroscopic information on how these ordered phases develop and evolve in response to perturbations - information that is not available from either static or time-resolved optical probes \cite{FiebigAPL,Tobey},  x-ray absorption spectroscopy (XAS) \cite{RiniPRB}, or transport measurements \cite{FiebigSci,RiniNat}. 

TR-RSXS research to date \cite{Durr,LSMO,CuO,LeeWS,Forst} has focused on the ultrafast disordering  or ``melting'' of electronic ordering within the first 100 ps following photo-excitation. However, the re-establishment of such ordering from a transient disordered state has been substantially overlooked. In this report, we focus on the re-establishment of electronic ordering from a metastable phase. Using TR-RSXS to follow the antiferromagnetic spin ordering (SO) in a transiently photo-excited Pr$_{0.7}$Ca$_{0.3}$MnO$_3$ \cite{TomiokaPRB96} over an unprecedented temporal window spanning 12 decades, we reveal the glass-like recovery dynamics with recovery times varying from sub-$\mu$s at low pump fluence to tens of seconds at high pump fluence. This is in striking contrast to the much faster response of charge carriers through electronic or lattice interactions \cite{Taylor}. Moreover, our data point to a crossover behavior in the dimensionality of the effective restoring interaction from quasi-1D at low pump fluence to 3D at high pump fluence.  Comparison to time-resolved resistivity measurements shows that although the collaspse of spin ordering is correlated with the decrease of resistivity, the much shorter recovery time for resistivity suggests that melting of spin ordering alone is not sufficient for inducing such a phase transition.

\C{\bf Negligible change in the in-plane correlation length upon melting}

TR-RSXS experiments were carried out at the ultrafast soft x-ray Beamline 6.0.2 of the Advanced Light Source (ALS), Lawrence Berkeley National Laboratory. Previous static RSXS studies showed that the (1/4, 1/4, 0) superlattice diffraction peak (in pseudocubic notation) at 65 K was dominated by the CE-type \cite{CE} SO state (see Fig.~1(b) for a characteristic RSXS energy profile) \cite{ZhouPRL}, and we focus on the dynamics of this state in the current study. An 800 nm pump laser pulse with 100 fs duration was used to induce an IMT \cite{FiebigSci}, and a 70 ps x-ray pulse was used subsequently to capture snapshots of the evolving SO. A schematic experimental geometry is shown in Fig.~1(a). The measurement temperature was 65 K, low enough to avoid any laser-heating induced thermal phase transition. 

\begin{figure}
\includegraphics[width=15 cm] {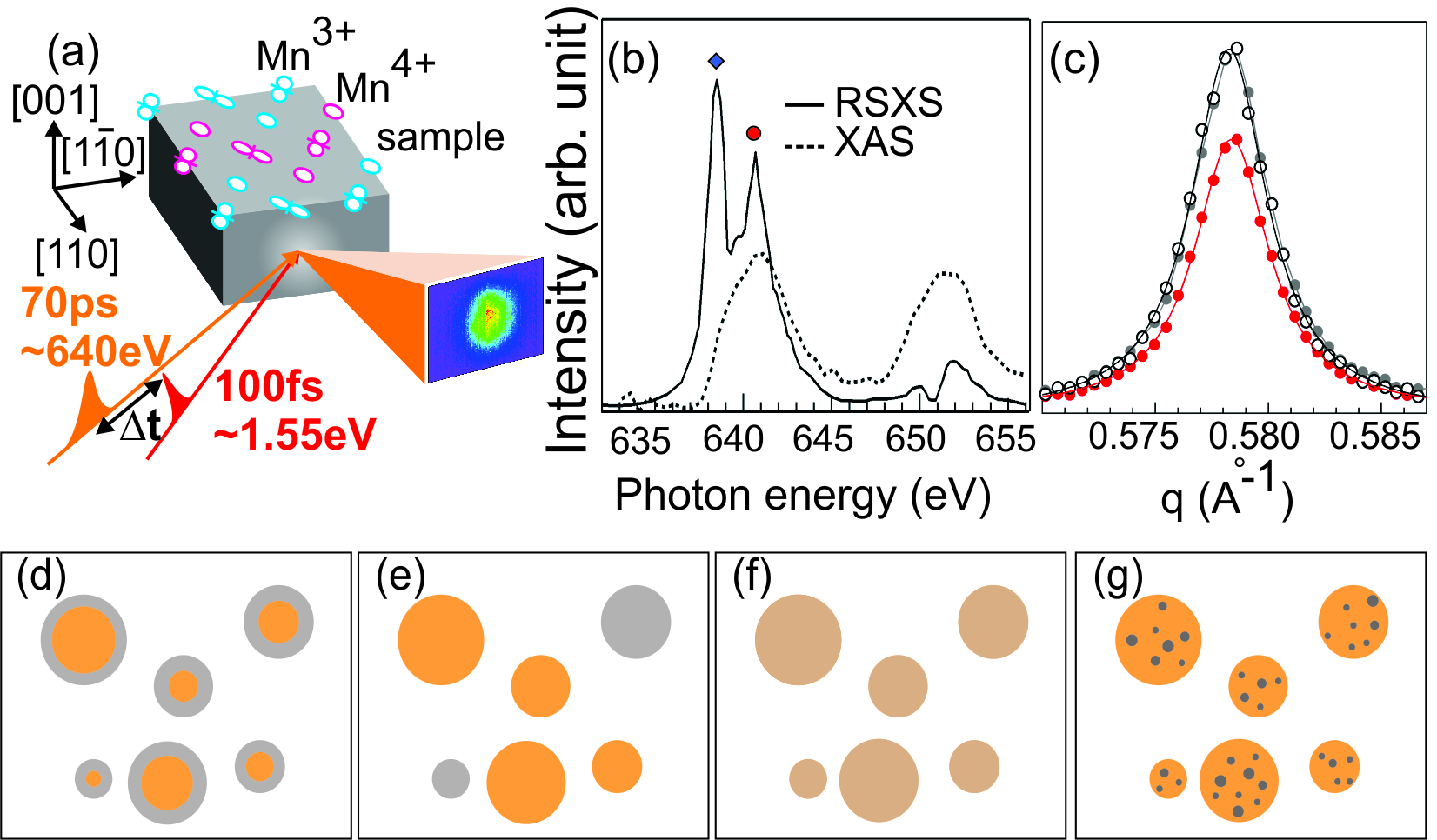}
\label{Figure 1}
\caption{(a) Schematic diagram of the CE-type SO state overlaid with the experimental geometry for TR-RSXS measurements. Circles and lobes represent the Mn$^{4+}$ sites and e$_g$ orbitals of the Mn$^{3+}$ sites. Pink and blue colors represent opposite spin orientations. (b) Energy profile of the SO diffraction peak from static RSXS measurements (solid line) as compared to the XAS spectrum (dotted line). The symbols mark the photon energies used in the time-resolved measurements. (c) Comparison of the SO diffraction peak intensity measured at 641.4 eV before (open circles) and $\Delta$t=500 ps after 800 nm laser excitation at 1 mJ/cm$^2$ pump fluence (red filled circles). Solid lines are fits using a Lorentzian function. Gray filled circles are data taken at $\Delta$t=500 ps rescaled by a factor of 1.33. (d-g) Schematic illustrations of four different responses of SO domains to photo-excitation. The orange areas represent the SO domains and the gray areas represent the regions destroyed by photo-excitation. These are over-simplified models and a realistic picture is likely more complicated due to the irregular shapes and interconnections between SO domains. }
\end{figure}

Figure 1(c) shows a comparison of the SO diffraction peak profiles in momentum space (q) before (open symbols) and at a delay time $\Delta$t=500 ps (red filled symbols) after photo-excitation (defined at t$_0$). The pump fluence is 1 mJ/cm$^2$ and the energy of probe x-ray beam is tuned to the Mn L$_3$ edge at 641.4 eV (red symbols in Fig.~1(b)). A reduction of the SO peak intensity is observed at $\Delta$t=500 ps. However, the peak position and width show negligible change upon photo-excitation, which can be clearly seen after rescaling the peak profile by the intensity ratio (filled gray symbols). A Lorentzian function fit to the data (solid lines in Fig.~1(c)) shows that the in-plane correlation length $\xi$, which is defined as $\xi$=2$\pi$/$\Delta$q where $\Delta$q is the peak width, remains 1560 $\AA$ $\pm$ 30 $\AA$ even following the suppression of the SO. While this study focuses on the in-plane (MnO$_2$) correlation length, it is possible that a more subtle change may be revealed in the out-of-plane correlation length if the three-dimensional scattering volume is measured. However, even in the case of La$_{0.5}$Sr$_{1.5}$MnO$_4$ where a  change is revealed \cite{CavalleriPRB2012}, the change in the out-of-plane correlation length is very subtle, on the order of a few percent or less. Considering that Pr$_{0.7}$Ca$_{0.3}$MnO$_3$ is more three-dimensional than the single-layered La$_{0.5}$Sr$_{1.5}$MnO$_4$, the change in the out-of-plane correlation length may be even more subtle. Such a  small change in the correlation length is still suprising considering the strong (and even complete) suppression of the SO peak intensity, and is in stark contrast to $\sim$ 10 times change in the  SO correlation length in the thermally induced phase transition across the canted-antiferromagnetic transition temperature T$_{CA}$ \cite{ZhouPRL}. 

The negligible change in the in-plane correlation length rules out the nucleation type recovery behavior illustrated in Fig.~1(d). Thus, there are three possible scenarios that may be consistent with the negligible change of correlation length: complete disordering of some SO domains without affecting other domains (Fig.~1(e)), phase fluctuations of the entire ordered domains (Fig.~1(f), homogeneous picture), or photo-induced local spin-disodered defects within the SO domains (Fig.~1(g), inhomogeneous picture).  These three scenarios can be further distinguished by their different dynamic responses to photo-excitation.

\C{\bf Glass-like recovery dynamics of spin ordering}

\begin{figure*}
\includegraphics[width=16.8 cm] {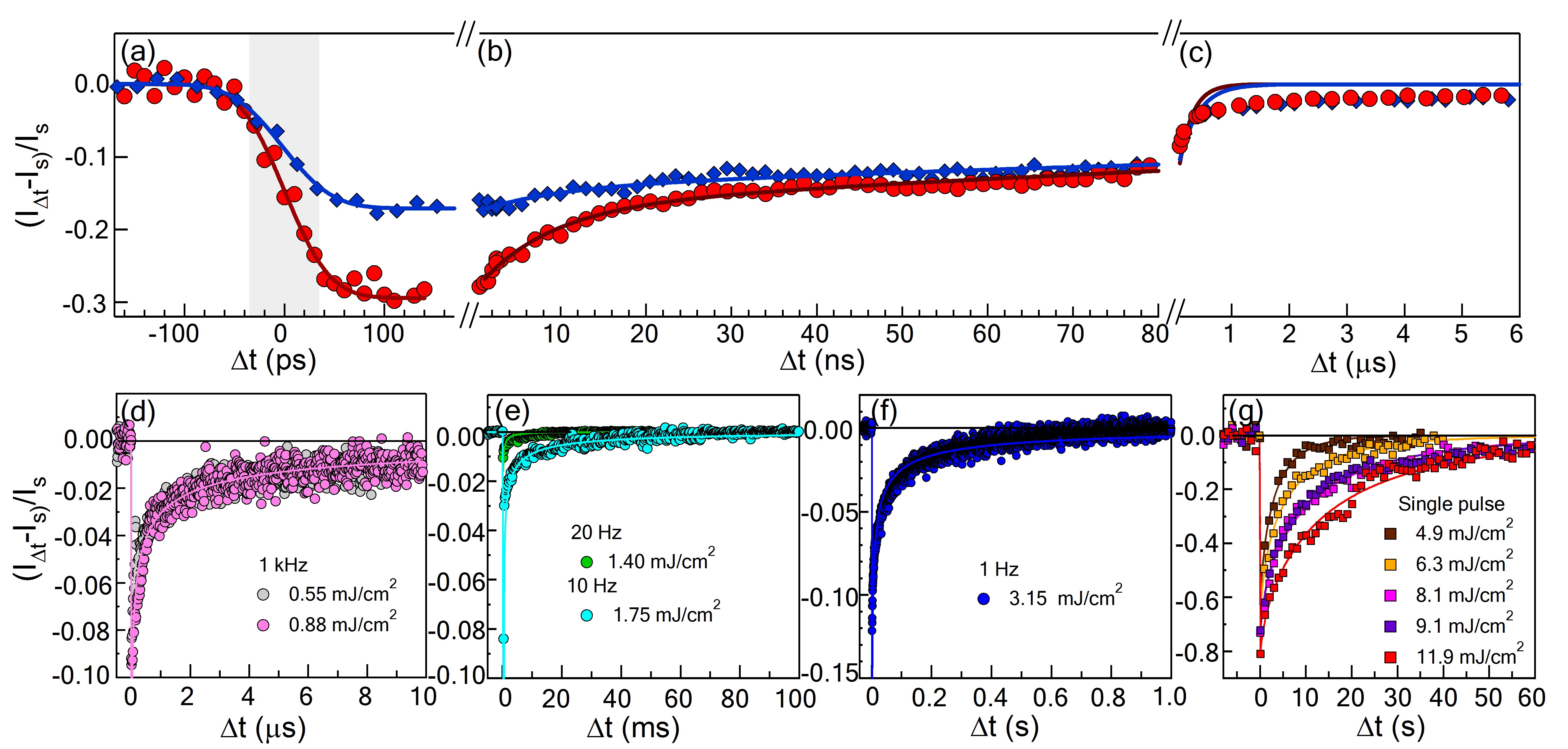}
\label{Figure 2}
\caption{(a-c) Differential SO peak intensity measured at 641.4 eV (red circles) and 639.2 eV (blue diamonds) as a function of pump-probe delay time $\Delta$t at 1 mJ/cm$^2$ pump fluence. The solid lines in panel (a) are fits using the error function and the shaded area marks the 70 ps temporal resolution of probe x-ray beam. The bi-exponential function fits up to 80 ns are shown as solid lines in panels (b) and (c). (d-g) Differential SO peak intensity as a function of delay time with different pump fluences. Symbols are raw data and solid lines are the stretched-exponential function fits. Data shown in panels (d-f) were taken with different laser repetition rates to ensure that the sample is recovered before the arrival of next pump laser pulse. The differential signals at very early delay times are off scale in panels (d-f) in order to show the recovery dynamics.  Data in panel (g) were measured with one single excitation pulse and the entire recovery process is measured with multiple x-ray pulses without re-exciting the sample (non-repetitive). The fitting parameters are listed in Table 1 of the supplementary information.}
\end{figure*}

Figure 2 shows the evolution of the SO diffraction peak intensity as a function of delay time $\Delta$t. The differential intensity is defined as $\Delta$I/I$_{s}$= (I$_{\Delta t}$-I$_s$)/I$_s$ where I$_s$ and I$_{\Delta t}$ are the peak intensity recorded before and at $\Delta$t after photo-excitation. A step-like decrease  is observed at $\Delta$t=0 with a fall time of 70 ps (Fig.~2(a)), which is limited by the pulse duration of the probe x-ray beam. Here we focus on the recovery dynamics at later delay times. 

Figures 2(b-c) show the re-establishment of SO in different temporal windows. Although a bi-exponential function a$_1$e$^{-t/\tau_1}$+a$_2$e$^{-t/\tau_2}$ with $\tau_1$ $\approx$ 10 ns and $\tau_2$ $\approx$ 100 ns seems to give a reasonable fit to the data within the first 80 ns (solid lines in Fig.~2(b)), a significant discrepancy is evident in the $\mu$s regime (Fig.~2(c)). In fact, the number of exponentials needed depends strongly on the selected temporal window, which is a clear indication of the inadequacy of using a multi-exponential function to describe the dynamics. The observed recovery behavior does not depend on the photon energy of the probe x-ray beam (red circles vs. blue diamonds in Figs.~2(a-c)), although different photon energies lead to variations in the probe depth \cite{ThomasPRL} and differential signal amplitude. The re-establishment of SO not only exhibits multiple time scales, but also strongly depends on the pump fluence. Figures 2(d-g) show the recovery of SO with increasing pump fluence. The full recovery time increases rapidly from sub-$\mu$s at low pump fluences (Fig.~2(d)) to tens of seconds at higher pump fluences (Fig.~2(g)). The surprisingly long recovery time rules out conventional electronic or lattice interactions as mediating mechanisms, as they typically occur on much shorter time scales (a few hundred ps or less) \cite{Taylor} . 

The strong fluence dependent recovery time scales resemble the dynamics seen in glass-like or complex disordered systems \cite{Phillips}, such as structural glass \cite{Glass}, magnetic glass \cite{Coey} etc. To substantiate this, we fit the data using the stretched-exponential function (Kohlrausch-Williams-Watt function) in the form of a$_0$e$^{-(t/\tau)^\beta}$ \cite{Phillips}, which has been used to describe glass-like dynamics.  Remarkably, this functional form nicely fits all data measured over 12 decades in time (see supplementary information for a detailed comparison between fits using a few commonly used non-exponential functions), and this strongly points to the glass-like nature of the process by which SO is re-established from a transient photo-excited state.  It is interesting to note that the recovery of the long-ranged SO domains shares similar dynamics to glass-like systems which are typically highly disordered. In contrast to Pr$_{0.7}$Ca$_{0.3}$MnO$_3$, TR-RSXS data from Pr$_{0.5}$Ca$_{0.5}$MnO$_3$,  which has a much more robust charge/orbital/spin ordered ground state at low temperature, does not show such clear glass-like behavior even at a pump fluence as high as 6 mJ/cm$^2$. The proximity of different competing ground states in Pr$_{0.7}$Ca$_{0.3}$MnO$_3$ and the associated frustration and phase separation are in agreement with the observed glass-like behavior, since frustration is an important ingredient for glass-like systems \cite{DagottoRP}. Although various types of glass-like behavior have been reported in manganites, e.g. spin glass \cite{SpinGlassManganite, Mathieu}, cluster glass \cite{ClusterGlass}, polaron glass \cite{ArgyriouPRL}, strain glass \cite{StrainGlass,WuNatMat} etc, our TR-RSXS work demonstrates that the long-ranged ($\xi$ $\approx$ 1500 $\AA$) SO phase after photo-excitation can manifest glass-like dynamics, and this finding is fundamentally different from the short ranged spin glass (negligible $\xi$) or polaron glass ( $\xi \le$ 10 nm) \cite{ArgyriouPRL, LynnPRB} previously discussed in manganites under equilibrium conditions.

The glass-like recovery dynamics can be further used to distinguish the three possible scenarios illustrated in Figs.~1(e-g). The scenario in Fig.~1(e) can be ruled out, since if individual SO domains are disordered and subsequently recover independently, the recovery time is not expected to show such strong pump fluence dependence. Although the phase fluctuation scenario (Fig.~1(f)) has been applied to explain the charge and spin stripe dynamics in a nickelate \cite{LeeWS},  Pr$_{0.7}$Ca$_{0.3}$MnO$_3$ in the current study  is very different. In contrast to the relatively simple 1D charge and spin stripes in the nickelate, Pr$_{0.7}$Ca$_{0.3}$MnO$_3$ exhibits CE-type zigzag spin ordering, with strong coupling to charge/orbital orderings and cooperative Jahn-Teller distortions, all of which give rise to numerous competing ground states and substantial inhomogeneity \cite{DagottoRP}. The large range of recovery times, and strong dependence of recovery time on pump fluence both point to a system with strong inhomogeneity and frustration.  In contrast, it is not clear how a more global  spin phase fluctuation model can explain these observations. Based on the negligible change of correlation length, the glass-like dynamics, and the corresponding fluence dependence of the conductivity (as discussed below), we propose a microscopic picture in which photo-excitation creates local metallic regions of frustrated spins within the larger SO domains (Fig.~1(g)). The spin disordered regions are sufficiently small and uncorrelated that they do not affect the overall SO domain size, or the pre-established correlation of the larger SO domains.  The variable sizes and frustrated spins that characterize these regions lead to multiple recovery time scales and the observed glass-like behavior. 

\C{\bf Pump fluence dependence of stretched-exponent $\beta$ and dimensional crossover}

\begin{figure*}
\includegraphics[width=16.8 cm] {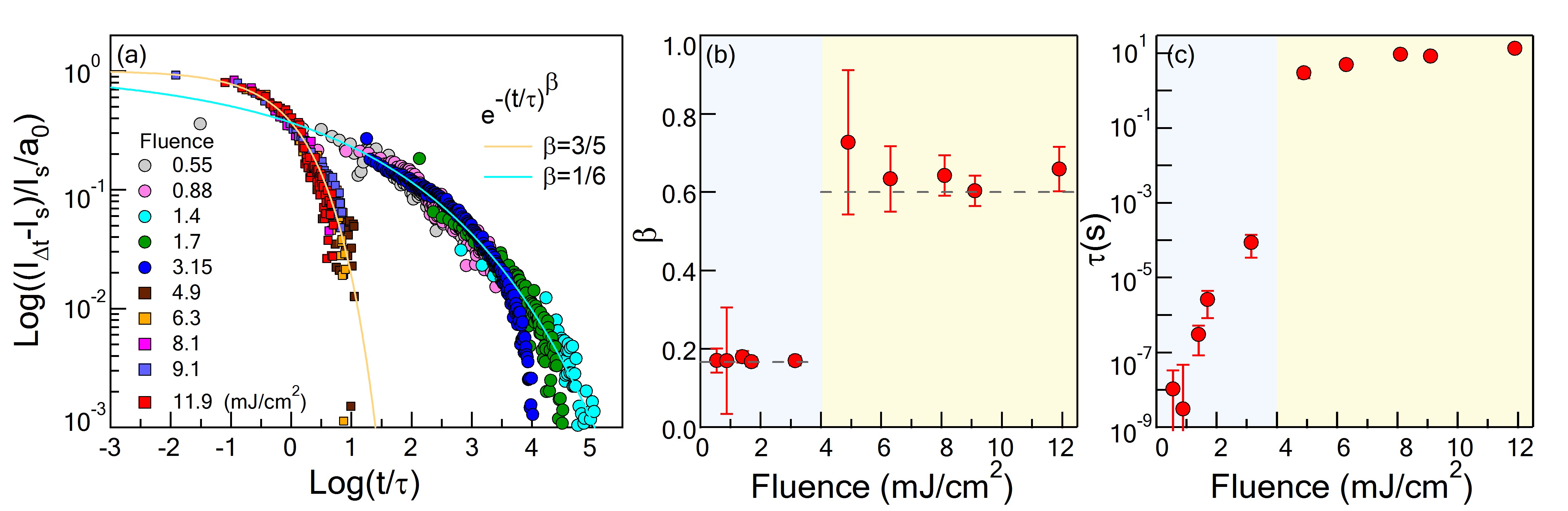}
\label{Fig3}
\caption{(a) Log-Log plot of the data shown in Figs.~2(d-g). The differential signal and time scale are normalized by a$_0$ and $\tau$ extracted from the stretched-exponential function fits. The cyan and yellow lines represent the stretched-exponential functions with $\beta$=1/6 and $\beta$=3/5. (b) The stretched exponent $\beta$ and error bars extracted from the fits as a function of pump fluence. (c) Extracted recovery time $\tau$ (on a Log scale) as a function of pump fluence.}
\end{figure*}

The effective dimensionality d of the interaction that re-establishes SO can be retrieved from the stretched-exponent $\beta$ via the relationship $\beta$= d/(d + 2) \cite{Phillips}.  Renormalizing the time traces in Figs.~2(d-g) by a$_0$ and $\tau$ shows a clear crossover behavior from $\beta$ $\approx$ 1/6 to $\beta$ $\approx$ 3/5 (d=3)  around 4 mJ/cm$^2$, and such evolution is also evident in the extracted $\beta$ (Fig. 3(b)). A crossover behavior is also observed in the pump fluence dependent recovery time (Fig.3(c)): below 4 mJ/cm$^2$ the recovery time increases by order of magnitude with pump fluence, while above 4 mJ/cm$^2$ the recovery time is saturated at tens of seconds. This pump fluence corrsponds to $\sim$ 10$\%$ in volume fraction (one photo-excitation per ten unit cells), assuming that the excitation is strictly confined to a single unit cell. However, we expect that the interaction may extend to neighboring sites and this fluence is within the range of percolation thresholds reported in manganites \cite{DagottoRP}. The ability to manipulate and sustain the metastable metallic phase \cite{FiebigSci} is likely a consequence of the glass-like nature of this phase.

 The quasi-1D structure of the zig-zag spin ordering may suggest a $\beta$ value close to 1/3 according to the continuous diffusion-to-trap model used to derive the $\beta$=d/(d+2) relationship \cite{PercoldationPRL2008, PhillipsarXiv}. However, it is known that $\beta$ can be strongly influenced by the microscopic details of the restoring interactions, and this may well explain the deviation from the 1/3 value predicted from the simplest model (or as observed in conventional spin glass) \cite{PercoldationPRL2008, PhillipsarXiv}. In particular, the recovery of the SO in Pr$_{0.7}$Ca$_{0.3}$MnO$_3$ is mediated by Goodenough-Kanamori rules \cite{Kanamori}.  The CE-type spin ordering is in a zig-zag pattern, with strong coupling to charge/orbital orderings and cooperative Jahn-Teller distortions.  In this sense, the SO in Pr$_{0.7}$Ca$_{0.3}$MnO$_3$ is substantially different from a 1D system with simple diffusion, and therefore it is not too surprising that the recovery is more restricted than 1D. At high pump fluences, the density of spin-disordered (and charge-disordered) defects is sufficiently high that inter-chain interactions become important and the restoring interaction is likely 3D.  We note that while dimensional crossover behavior has been reported in many systems by tuning thermodynamic variables under equilibrium conditions \cite{Valla}, our results represent its manifestation in a dynamic regime. 

\C{\bf Role of melting of spin ordering in the photo-induced IMT}

\begin{figure*}
\includegraphics[width=16.8 cm] {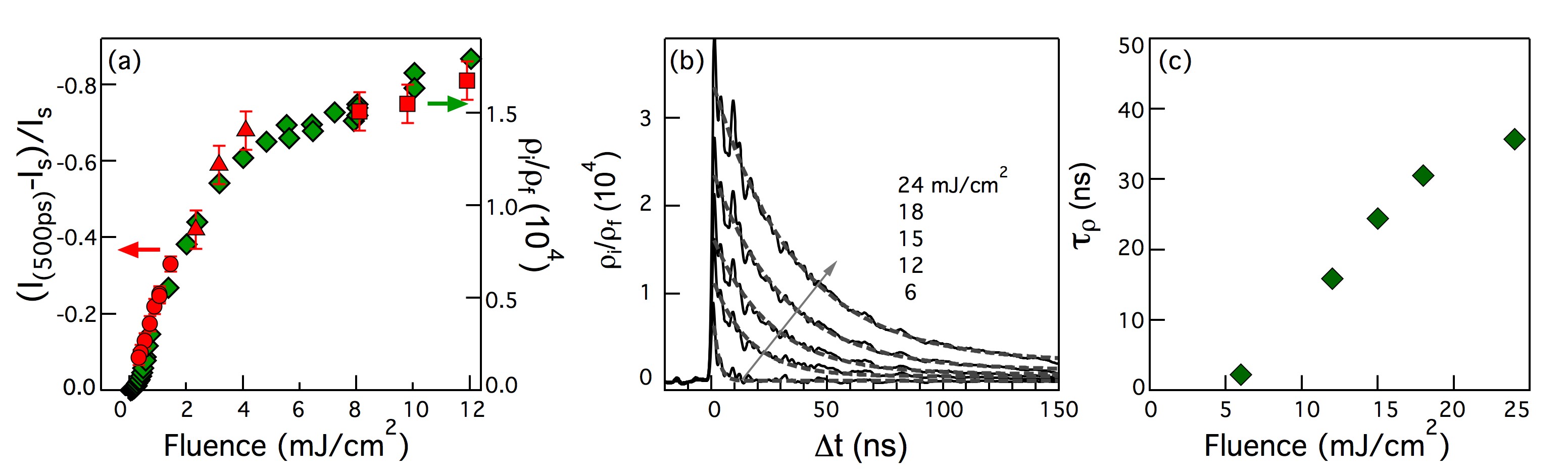}
\label{Fig3}
\caption{Comparison of TR-RSXS results with time-resolved resistivity measurements. (a) Comparison between the maximum transient differential SO peak intensity (red symbols, left axis) and maximum resistivity change (green diamonds, right axis) right after photo-excitation as a function of pump fluence. Circles, triangles and squares were taken at $\Delta$t= 500 ps, 0.5 $\mu$s and 0.1 s respectively. (b) Transient resistivity change $\rho_{i}/\rho_{f}$ as a function of delay time. Solid lines are the data and dotted lines are the fits using a single exponential function. The oscillations at early delay time are due to electronic ringings and are not real response from the sample.  (c) Extracted resistivity recovery time $\tau_{\rho}$ as a function of pump fluence. }
\end{figure*}

A connection between melting of the SO and the photo-induced IMT can be established by comparing the results from TR-RSXS and time-resolved resistivity measurements. Figure 4(a) shows a comparison of the differential SO scattering intensity $\Delta$I/I$_S$ (red symbols) and the maximum change in resistivity ${\rho_i/\rho_f}$ (green diamonds, $\rho_i$ and $\rho_f$ are the initial and transient resistivity right after t$_0$). Since the resistivity change gauges the local metallicity, which forms a macroscopic conducting path when stabilized by an external voltage \cite{FiebigSci}, the strong similarity in the pump fluence dependence indicates that photo-carriers liberated from the SO state actively participate in producing local metallic domains, i.e. the collapse of the SO state is directly correlated with the photo-induced IMT.

Interestingly, while our results show that the establishment of metallicity is commensurate with the melting of SO, the recovery of SO does not coincide with the re-establishment of the insulating state.  Figure 4(b) shows the time-dependent resistivity change, which can be used to extract the resistivity recovery time $\tau_{\rho}$. Unlike TR-RSXS data, a single exponential function is sufficient to fit the time-resolved resistivity data, suggesting very different dynamics between spin ordering and resistivity. Although the recovery time scales for the SO state $\tau$, plotted on a Log scale in Fig~3(c), increase rapidly with pump fluence by nearly 9 orders of magnitude from tens of ns to tens of seconds, the resistivity recovery time $\tau_{\rho}$ (plotted on a linear scale in Fig.4(c)) remains below 50 ns even at a fluence of 24 mJ/cm$^2$. Two factors may account for the faster recovery time for resistivity compared to the slow re-establishment of SO.  First, manganites exhibit nanoscale phase separation of insulating and metallic domains \cite{DagottoRP}.  In such a phase separation or percolation scenario, the insulating state will recover as soon as the continuous metallic pathway is disrupted.  In contrast, the re-establishment of spin ordering will continue within isolated metallic domains, as is evident in the recovering TR-RSXS signal, without influencing the macroscopic resistivity.  Second, the disruption of SO is likely concurrent with the disruption of CO and with the establishment of metallic domains.  However, CO may be re-established on faster time scales (and possibly without the glass-like dynamics associated with frustrated SO), leading to a CO insulating phase with residual frustrated disorder of the spins. Even though time-resolved resistivity and TR-RSXS are two experimental probes involving different electronic degrees of freedom, length scales and physical properties, the complimentary information provided by these two probes can yield a much deeper understanding on the underlying physics of the complex phenomena.

\C{\bf Summary}

TR-RSXS studies of SO dynamics in Pr$_{0.7}$Ca$_{0.3}$MnO$_3$ covering 12 decades in time reveal that the recovery of SO from a transiently photo-excited state manifests glass-like behavior with a dimensional crossover in the effective interaction from quasi-1D at low pump fluence to 3D  at high pump fluence.  Comparison of SO dynamics with time-resolved resistivity measurements suggests that the collapse of spin ordering is correlated with the IMT.  However, the re-establishment of SO is not necessary for the recovery of the insulating phase.  This may be a consequence of percolation and/or the establishment of a novel CO (insulating) phase containing residual disordered and frustrated spins. Our work provides a new perspective for revealing the fascinating physics hidden in the recovery dynamics of electronic ordering in correlated electron materials after transient photo-excitation, a prominent method for ultrafast manipulation of material properties. Since various transition metal oxides that exhibit intriguing emergent phenomena, e.g. cuprates, have rich competing phases that involve dynamic electronic orderings, it will be important to extend similar studies to those systems. For example, charge stripes \cite{Tranquada} are reported in cuprates and studies of the stripe dynamics associated with the photo-induced superconductivity \cite{CavalleriSci} may shed new light on the mechanism of high temperature superconductivity.

{\bf Methods}

Single crystal Pr$_{0.7}$Ca$_{0.3}$MnO$_3$ samples were grown by traveling solvent floating-zone method. The samples were first oriented using the Laue diffraction pattern and then cut and polished to produce an optically flat [110] surface. A systematic study of spin and orbital ordering (SO/OO), as a function of temperature, sample orientation, and hole doping, is presented in Ref. \onlinecite{ZhouPRL}. 

In TR-RSXS experiments, the sample was oriented with c-axis perpendicular to the scattering plane to enhance the sensitivity to spin ordering \cite{ZhouPRL}. At 641.4 eV, the sample and detector angles ($\theta$ and 2$\theta$) were $\approx$ 61$^\circ$ and 122$^\circ$. An amplified laser system, with tunable repetition rate from 4 kHz to 1 Hz, was used for pumping the sample. In all the repetitive measurements, we have ensured that the diffraction signal is recovered between pulses  by comparing the static diffraction signal I$_s$ (i.e. before applying any laser excitation) with the transient diffraction signal at negative delay: I$_{\Delta t}$ with $\Delta$t=0$^-$. The repetition rate is chosen to insure that this residual signal is small (I$_s$-I$_{\Delta t = 0^- }$)/I$_s$  $\le$ 2$\%$).  In addition, the negligible effect of the small residual signal on the fitting parameters is confirmed using the model detailed in the supplementary information. 

In repetitive measurements, the laser repetition rate was intentionally reduced at higher pump fluence to ensure that the SO signal is recovered before the arrival of the next laser pulse. The pump laser beam was introduced at $\approx$ 15$^\circ$ relative to the incident x-ray beam (closer to sample normal) to avoid direct reflection of the laser beam into the photon detector. An avalanche photodiode (APD), enclosed inside an aluminum box with a high quality (pinhole-less) 200 nm thick aluminum window in the front to filter out ambient light, was used to recorded the diffracted spin ordering signal. A voltage amplifier was used to amplify the APD signal before sending it to a boxcar or fast oscilloscope for data acquisition. The TR-RSXS signal does not depend on the polarization of the pump laser beam, which was kept in the horizontal scattering plane for data shown here.

Time-resolved resistivity measurements were performed at 70 K by measuring the voltage drop across a 50 $\Omega$ reference resistance in series with the Pr$_{0.7}$Ca$_{0.3}$MnO$_3$ samples using a 1 GHz oscilloscope, while shining a laser pulse onto the gap of 150-200 $\mu$m between two gold electrodes and applying a bias voltage of 5-30 V.

\begin{acknowledgments}
We thank D.-H. Lee, E. Dagotto, J. Orenstein, R.A. Kaindl, H. Yao and W.L. Yang for useful discussions. This work was supported by the Director, Office of Science, Office of Basic Energy Sciences, the Materials Sciences and Engineering Division under the Department of Energy Contract No. DE-AC02-05CH11231. The Advanced Light Source is supported by the Director, Office of Science, Office of Basic Energy Sciences, of the U.S. Department of Energy under Contract No. DE-AC02-05CH11231.
\end{acknowledgments}

{\bf Additional Information}

{\bf Author Contributions}

R.W.S, Z.H. and S.Y.Z. conceived the tr-RSXS project. S.Y.Z, Y.Z. and M.C.L. performed the tr-RSXS experiments. M.R. performed the time-resolved resistivity measurements. T.E.G and M.P.H provided support for the ultrafast X-ray beamline. A.G.C.G, N.T. and Y.-D.C. provided support for RSXS experiments. Y. Tomioka and Y.Tokura provided the single crystal samples.  S.Y.Z. and M.C.L. analyzed tr-RSXS data. S.Y.Z., Y.-D.C. and R.W.S. wrote up the manuscript.

{\bf Competing Financial Interests}

The authors declare no competing financial interests.

\end{document}